\documentclass[12pt]{article}
\usepackage{amsmath,amssymb}

\usepackage{tikz}
\usetikzlibrary{decorations.pathreplacing,decorations.markings,decorations.text}
\usetikzlibrary{arrows,shapes.geometric,calc}
\usepackage{mathtools}

\definecolor{BB}{RGB}{128,184,220}

\usepackage{graphicx}
\usepackage{wrapfig}

\usepackage{hyperref}
\usepackage{cite}

\def\beq{\begin{eqnarray}}
\def\eeq{\end{eqnarray}}

\def\la{\langle }
\def\ra{\rangle }

\newcommand{\Tr}{\,\mathrm{Tr}\,}            

\newcommand{\CO}{{\mathcal{O}}}
\newcommand{\CH}{{\mathcal{H}}}

\newcommand{\be}{\begin{equation}}
\newcommand{\ee}{\end{equation}}
\newcommand{\bea}{\begin{eqnarray}}
\newcommand{\eea}{\end{eqnarray}}
\newcommand{\bg}{\begin{gather}}

\newcommand{\bseq}{\begin{subequations}}
\newcommand{\eseq}{\end{subequations}}

\renewcommand{\tanh}{\mathop{\rm th}\nolimits}

\renewcommand{\ln}{\mathop{\rm ln}\nolimits}

\def\tr{\hbox{Tr}}

\def\be{\begin{eqnarray}}
\def\ee{\end{eqnarray}}
\def\lb{\label}

\newcommand{\CM}{{\mathcal{M}}}

\newcommand{\CP}{{\mathcal{P}}}

\newcommand{\CS}{{\mathcal{S}}}

\newcommand{\p}{\partial}
\begin{document}

\title{\textbf{Holographic calculation of entanglement \\ entropy 
in the presence of boundaries }}

\date{}

\vspace{1.cm}
\author{ \textbf{  
{\rm \normalsize Amin Faraji Astaneh$^{1}$, Cl\'ement Berthiere$^2$, 
Dmitri Fursaev$^3$ and  Sergey N. Solodukhin$^2$}
 }} 
\maketitle
\begin{center}
\hspace{-0mm}
  \emph{$^1$ School of Particles and Accelerators,\\ Institute for Research in Fundamental Sciences (IPM),}\\
\emph{ P.O. Box 19395-5531, Tehran, Iran}
 \end{center}
\begin{center}
  \hspace{-0mm}
  \emph{ $^2$ Laboratoire de Math\'ematiques et Physique Th\'eorique  CNRS-UMR
7350, }\\
  \emph{F\'ed\'eration Denis Poisson, Universit\'e Fran\c cois-Rabelais Tours,  }\\
  \emph{Parc de Grandmont, 37200 Tours, France}
\end{center}
\begin{center}
\hspace{-0mm}
  \emph{$^3$ Dubna State University\\ Universitetskaya str.~19,
  141 980 Dubna, Moscow Region Russia,}\\
\emph{and the Bogoliubov Laboratory of Theoretical Physics, JINR, Dubna}
 \end{center}



\vspace{0.4cm}

\begin{abstract}
\noindent {When a spacetime has boundaries, the entangling surface does not have to be necessarily compact and it may have boundaries as well.
Then there appear a new, boundary,  contribution to the entanglement entropy due to the intersection of the entangling surface with the boundary of the spacetime.
We study the boundary contribution to the logarithmic term in the entanglement entropy in dimensions $d=3$ and $d=4$ when the entangling surface is orthogonal to the
boundary. In particular,  we compute a  boundary term in the entropy of
${\cal N}=4$ super-gauge multiplet at weak coupling. For gauge fields 
we use a prescription which is consistent with the positive 
area law.
The boundary term is compared with the holographic calculation of the entropy based on the Ryu-Takayanagi proposal adapted appropriately  to the present situation.  We find a complete agreement between these two calculations
provided the boundary conditions imposed on the gauge multiplet
preserve $1/2$ of the supersymmetry and the extension of the boundary into the AdS bulk is a minimal hypersurface.
}
\end{abstract}

\vskip 0.4 cm
\noindent
\rule{7.7 cm}{.5 pt}\\
\noindent 
\noindent
\noindent ~~~ {\footnotesize e-mails:\ faraji@ipm.ir ,  clement.berthiere@lmpt.univ-tours.fr , fursaev@theor.jinr.ru , \\
Sergey.Solodukhin@lmpt.univ-tours.fr}


\newpage

\section{ Introduction}
Entanglement entropy has found many interesting applications from the physics of black holes to 
condensed matter and quantum computers. The intriguing property of entanglement entropy is
that it is closely related to geometry. To leading order in UV cut-off it is proportional to the area of 
the entangling surface while the subleading terms, such as logarithmic, contain information about the topology of the surface and on the way it is embedded in spacetime.  

A new aspect of the relation between the entropy and geometry appears when the spacetime ${\cal M}_d$ in question has a boundary, $\partial{\cal M}_d$.
Then the entangling surface $\Sigma$ may end on the boundary and, thus, may have some boundary itself, $\partial\Sigma={\cal P}$ 
(${\cal P}=\partial{\cal M}_d\cap \Sigma$). In this situation there may appear new terms in the entanglement entropy that are due to the intersection
${\cal P}$ and that are absent if spacetime does not have boundaries. This new aspect has been studied in a holographic framework \cite{Takayanagi:2011zk}
and on the field theory side \cite{Fursaev:2006ng},  \cite{Fursaev:2013mxa},  \cite{Fursaev:2016inw}, \cite{C-S}.  The most interesting terms in the 
entropy are logarithmic. They are considered to be independent of the regularization. Moreover, for the conformal field theories
the logarithmic terms are believed to contain information on the conformal anomalies,  as suggested in \cite{Solodukhin:2008dh}.

In the absence of boundaries, when $\partial \Sigma=\emptyset$, the logarithmic term in the entropy is constructed from the Riemann curvature of the spacetime and the extrinsic curvature 
of the entangling surface $\Sigma$. Since there are two normal vectors to $\Sigma$, there should necessarily be an even number of components of the
extrinsic curvature in such an invariant, so that the logarithmic term is vanishing if the dimension of spacetime is odd.
This goes in parallel with the fact that the local conformal anomaly is vanishing in odd dimensions. 

The story is quite different if there are boundaries, \cite{Herzog:2015ioa},  \cite{Fursaev:2015wpa},  \cite{Solodukhin:2015eca}, \cite{AW}. Then, the extrinsic curvature of the boundary may come into play in a way that it becomes possible to
construct invariants of odd dimensions. As a result, the integral  conformal anomaly can be non-vanishing even if the dimension of the spacetime is odd, see \cite{Solodukhin:2015eca}.
Respectively, there may appear new contributions to the logarithmic term which are non-vanishing due to the extrinsic curvature of the boundary $\partial{\cal M}_d$ localized at $\cal P$. The exact form of these terms 
in $d=4$ was suggested in \cite{Fursaev:2013mxa}. A related story for defects has been studied in \cite{Jensen:2015swa}.

When formulating a holographic approach to this sort of calculations the first question is how the boundary $\partial{\cal M}_d$ is extended into the bulk of the $(d+1)$-dimensional 
Anti-de Sitter spacetime. Suppose that ${\cal S}_d$ is such an extension, $\partial{\cal S}_d=\partial{\cal M}_d$. Historically, the first attempt to describe holographically the boundary conformal field theory was made in \cite{Takayanagi:2011zk}. In that approach  the surface  ${\cal S}_d$ is supposed to satisfy certain tensor-type equations 
obtained by varying the gravitational action with respect to metric induced on ${\cal S}_d$. These equations are so restrictive that the solution does not always exist.  
In order to define consistently the surface ${\cal S}_d$ for a generic boundary $\partial{\cal M}_d$, the equations should be reduced to a scalar-type equation on the trace of the extrinsic curvature of ${\cal S}_d$  \cite{SJ}. A related prescription was  considered in \cite{Miao:2017gyt}.

Recently, two of the present authors \cite{Astaneh:2017ghi}, have performed a holographic calculation of the boundary terms in the conformal anomaly in dimensions $d=3$ and $d=4$
and compared the result with the explicit field theory calculation for the ${\cal N}=4$ super-gauge multiplet.  The two calculations give rise to identical results
provided ${\cal S}_d$ is a minimal surface, that is the trace of its extrinsic curvature vanishes. Interestingly, the agreement takes place if on the field theory side
the boundary conditions imposed on the fields of the super-multiplet preserve $1/2$ of the supersymmetry \cite{Gaiotto:2008sa}.

In the present paper we extend this consideration to a holographic calculation of entanglement entropy. Our goal is two-fold. First, we want to compute the entropy for the free field  ${\cal N}=4$ super-multiplet and identify the conformal charge which corresponds to a new boundary term
in the entanglement entropy.  
Then, we perform the calculation holographically and compare the results.  Our second goal is to see whether the agreement between the field theory 
and the holographic calculations in the case of anomalies can be extended to the entanglement entropy.
The conclusion we make in this paper is that, quite remarkably,  the agreement works  perfectly both for the anomaly and the entropy.

The paper is organized as follows. In the next section we first review the field theory computation of the boundary term in the entanglement entropy in dimension $d=3$ and then do the direct calculation in dimension $d=4$, following the earlier paper 
\cite{Fursaev:2013mxa}, in the case of free fields of spin $s=0,~1/2~$ and $1$ and then for the entire ${\cal N}=4$ super-multiplet. Special attention is paid to the gauge sector where
calculation should be consistent with the positive leading area type contribution to the entropy.
In section 3 we develop a holographic approach to the same calculation and compare the results.

In this paper we consider a somewhat simplified situation when the entangling surface $\Sigma$  and the boundary $\partial{\cal M}_d$ intersect orthogonally.
In more general situations the entropy would be a function of the angles between the vectors normal to $\Sigma$ and  the normal vector to $\partial {\cal M}_d$.
Throughout the paper the normal vector to the boundary, $N^{\mu}$, is always outward and the extrinsic curvature is defined with respect to this outward normal vector.

\section{Calculation in field theory}

\subsection{A summary of calculation in $d=3$}

As we have discussed this in the Introduction, in a  odd-dimensional conformal field theory, there is no logarithmic term in the entanglement entropy and, in parallel, the local conformal anomaly is vanishing. However, if the spacetime has boundaries, this statement is no longer true. Indeed, on the boundary, which in this case is even-dimensional, one can construct certain invariants of appropriate dimensions using the extrinsic curvature  of the boundary. As a consequence, these boundary terms produce a non-trivial anomaly if the dimension of spacetime is odd.  Similarly, it was shown in \cite{Fursaev:2016inw} that the logarithmic term in the entanglement entropy is also non-trivial in dimension $d=3$, provided the entangling surface crosses the boundary of the spacetime. We briefly review the findings of \cite{Fursaev:2016inw}.

In three dimensions, the entanglement entropy is expected to have the following assymptotics:
\be
S(\Sigma) = \alpha\frac{L}{\epsilon} + s_{log}\ln\frac{\epsilon}{\mu} + \cdots\,,  
\lb{slog3d}
\ee
where $L$ is the length of $\Sigma$, $\epsilon$ is a UV cut-off and $\mu$ is a typical scale of the theory.
The logarithmic part of the entropy in (\ref{slog3d}) can be derived from the divergent part of the effective action on a manifold with conical singularities, using the heat kernel technology (we describe the method in section \ref{4dsec}). In general, it depends on the number of points where the entangling surface $\Sigma$ crosses the boundary, as well as the angles of those intersections. Thus, in three dimensions what we call $\cal P$ is a collection of such points.

If one chooses a situation where $\Sigma$ crosses orthogonally the boundary $\partial{\cal{M}}_3$ which consists of one or two parallel plane(s), there will be $n_p = 1\;{\rm or}\;2$ point(s) of intersection(s). Then one can show that the logarithmic term in the entanglement entropy is
\be
s_{log} = \frac{a}{24}n_p\,,
\lb{slog3}
\ee
where the charge $a$ is found in Table \ref{tab3} for scalars and Dirac spinors.

It is interesting to note that one can alternatively derive (\ref{slog3}) from the conformal anomaly of the effective action, 
\be
\int_{{\cal M}_3}\la T\ra= -\frac{a}{384\pi}\int_{\partial{\cal{M}}_3} \hat{R} + \frac{q}{256\pi}\int_{\partial{\cal{M}}_3} \Tr \hat{k}^2\,,
\lb{A3}
\ee
where $ \hat{R}$ is the Ricci curvature on the boundary and $\hat{k}_{\mu\nu}$ is the traceless part of the extrinsic curvature tensor of $\partial{\cal{M}}_3$.
The logarithmic term in the entropy computed directly from the anomaly (\ref{A3}) reads
\be
s_{anom} &=& s_{log} +s_{nc}\,, \lb{sanom}\\[1.2ex]
&& \hspace{-2.8cm} s_{anom} = \frac{a}{96}n_p\,,  \quad s_{nc}= - \frac{a}{32}n_p \, .\nonumber 
\ee
The term $s_{nc}$ comes from the non-minimal coupling with the curvature in the Laplacian operator for scalars and should be subtracted to $s_{anom}$ to recover $s_{log}$.
We refer the reader to \cite{Fursaev:2016inw} for further details\footnote{We use opposite signs conventions to \cite{Fursaev:2016inw} for $s_{log},\, s_{anom},\, s_{nc}$.}.

\begin{table}[h]\renewcommand{\arraystretch}{1.5}
\begin{center}
\begin{tabular}{|c|c|c|c|}
\hline
 Theory  & $a$ & $q$ & boundary condition  \\ 
\hline
 real scalar & $1$ & $1$ & Dirichlet   \\
 \hline
 real scalar & -$1$ & $1$ &  Robin   \\
 \hline
Dirac spinor & $0$ & $2$ & mixed   \\
\hline
\end{tabular}
\vspace{-10pt}
\end{center}
\caption{Charges in the anomaly of the effective action and $s_{log}$ in $d=3$ dimensions}
\label{tab3}
\end{table}

We see that the conformal charge $q$ in (\ref{A3}) does not produce any contribution to the entanglement entropy. This however may change if the angle between the 
entangling surface $\Sigma$ and the boundary $\partial{\cal M}_3$ is different than $\pi/2$. The discussion of the effects of the angle is beyond the scope of the present paper.

\subsection{Calculation in $d=4$}
\lb{4dsec}
\subsubsection{General structure of logarithmic term in entanglement entropy}
Before discussing the entropy we remind the general form for the integral conformal anomaly in four dimensions,
\be
\int_{{\cal M}_4}\la T\ra=-\frac{a}{180}\chi[{\cal M}_4]+\frac{b}{1920\pi^2}\left(\int_{{\cal M}_4} \;\;\,\mathclap{\Tr W^2} \quad\;\; -8\int_{\partial{\cal M}_4}W^{\mu\nu\alpha\beta}N_\mu N_\beta \hat{k}_{\nu\alpha}\right)+\frac{c}{280\pi^2}\int_{\partial{\cal M}_4}\;\mathclap{ \Tr\hat{k}^3}\quad\, ,
\lb{d4}
\ee
valid for free fields of spin $s=0\, , \  1/2\, , \ 1$, see \cite{Fursaev:2015wpa}. $a$ and $b$ are the well-known conformal charges that appear already in the local anomaly
while $c$ is a new charge  characterizing the part of the anomaly entirely due to the presence of the boundary. For completeness, the values of $a\, , \ b\, , \ c$ are given below in Table 2.

\medskip

In four dimensions, the entanglement entropy has the following asymptotic 
dependence on the UV cut-off $\epsilon$,
\be
S(\Sigma) = \frac{s_2}{\epsilon^2} + \frac{s_1}{\epsilon} + s_{log}\ln\frac{\epsilon}{\mu} + \cdots\,,  
\ee
where $\mu$ is a typical scale of the theory. The leading term $s_2$ is proportional, as it is well-known, to the area of $\Sigma$, while $s_1$ is proportional to the length of $\cal P$, the intersection of $\Sigma$ with $\partial\mathcal{M}_4$ (see for example \cite{Fursaev:2006ng}, \cite{C-S}). 
The logarithmic term $s_{log}$ is a combination of conformal  invariants constructed on  $\Sigma$ and its boundary ${\cal P}=\partial\Sigma$\footnote{Let us note that charges
in (\ref{d4}), (\ref{slog}) are related 
to the corresponding charges $a'$,$c'$,$q_2'$,$d'$ in \cite{Fursaev:2013mxa}\cite{Fursaev:2015wpa} as follows: $a=360 a'$,$b=120 c'$,
$c=35q_2'/2$, $d=60d'$.},
\be
s_{log}\; =\; \frac{a}{720\pi}\left[\int_\Sigma R_\Sigma + 2 \int_{\cal P} k_p  \right] \hspace{-7pt}&-&\hspace{-7pt} \frac{b}{240\pi}\int_\Sigma [W_{ijij}  -\Tr \hat{k}_i^2] \;+\; d\, F_d \;+\; e\, F_e\,,\lb{slog}   \\ [1ex]
F_d =  -\frac{1}{40\pi}\int_{\cal P} \hat{k}_{\mu\nu}v^\mu v^\nu\,, && F_e = -\frac{1}{\pi}\int_{\cal P} (N\cdot p_i)(\hat{k}_i)_{\mu\nu}v^\mu v^\nu\,.\nonumber
\ee
The first two terms in (\ref{slog}) were found in \cite{Solodukhin:2008dh}, $a$ and $b$ are the  charges that appear in the local conformal anomaly and in the integral anomaly as in (\ref{d4}).
The first term, due to the $a$ charge, is the Euler number of the entangling surface. It has a boundary contribution due to ${\cal P}=\partial \Sigma$, $k_p$ is the extrinsic curvature
of ${\cal P}$.  In the second term $W_{ijij} = W_{\mu\nu\alpha\beta}p_i^\mu p_j^\nu p_i^\alpha p_j^\beta$ is the projection of the Weyl tensor on the subspace orthogonal to $\Sigma$,  $\{p_i^\mu \}$, $i=1,2$, are two unit mutually orthogonal normal vectors to $\Sigma$ and we define 
$ (\hat{k}_i)_{\mu\nu} = (k_i)_{\mu\nu} -\frac{1}{2}h_{\mu\nu}k_i$, $(k_i)_{\mu\nu}$ is the extrinsic curvature of $\Sigma$ as embedded in the four-dimensional spacetime, and $h_{\mu\nu}$ is the induced metric on $\Sigma$.

The last two conformal invariants in (\ref{slog}) were introduced in \cite{Fursaev:2013mxa}.  
Here $\hat{k}_{\mu\nu}=k_{\mu\nu}-\frac{1}{3}\gamma_{\mu\nu}k$, where
$k_{\mu\nu}$ is the extrinsic curvature of $\partial {\cal M}_4$, $\gamma_{\mu\nu}$ is the induced metric on $\partial {\cal M}_4$ and $v^\mu$ is a unit vector tangent to $\cal P$.
$F_d$ and $F_e$ are simplest invariant structures, $F_e$ reflecting
properties of extrinsic geometry of $\Sigma$ at the boundary and $F_d$ keeping the information about the extrinsic geometry of the boundary itself. 

Note that the expressions for $F_d$ and $F_e$  in (\ref{slog}) are not uniquely fixed by the conformal invariance requirements. They may depend on a tilt of $\Sigma$ to the boundary. Let $l$ be a unit outward looking vector at ${\cal P}$ which lies in a tangent space to $\Sigma$
and is orthogonal to ${\cal P}$. The tilt angle $\alpha$ at each point of ${\cal P}$ 
can be defined as $\cos\alpha=N_\mu l^\mu$. One can generalize $F_d$ and $F_e$ by including
there functions of $\alpha$ as pre-factors. One can also add to $F_d$ an extra term 
of the form $\hat{k}_{\mu\nu}l^\mu l^\nu$.

These ambiguities should be fixed by additional arguments which will not be discussed here.
In our work $\Sigma$ is assumed to cross $\partial{\cal M}_4$ orthogonally. 
In this case $l=N$, $F_e$ and $\hat{k}_{\mu\nu}l^\mu l^\nu$ vanish, while possible
constant prefactors in $F_d$ can be eliminated by redefinition of $d$.
Keeping this in mind in what follows we ignore $F_e$ and take for $F_d$ the above definition.

\medskip

Below  we shall determine the new charge  $d$, following the method suggested in  \cite{Fursaev:2013mxa}.

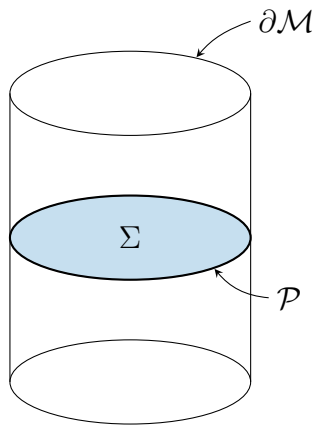
\begin{figure}[h]
\vspace{15pt}
\begin{center}
\begin{tikzpicture}[x=1.6cm,y=1.6cm]
	\draw[] (0,1.2) circle(1 and 0.35);
	\draw[] (0,-1.2) circle(1 and 0.35);
	\draw (-1,-1.2)--(-1.,1.2);
	\draw (1,-1.2)--(1,1.2);
	\draw[fill=BB,opacity=0.45] (0,0) circle(1 and 0.35);
	\draw[thick] (0,0) circle(1 and 0.35);
	\node at (0,0) {$\Sigma$};
	\node at (1.3,1.8) {$\partial\cal{M}$};
	\node at (1.32,-0.5) {$\cal{P}$};
	\node at (-1.15,-0.5) {};
	\draw[black,->,>=stealth] (1.,1.8) to[out=200, in=50, looseness=1] (0.55,1.5);
	\draw[black,->,>=stealth] (1.15,-0.5) to[out=175, in=-50, looseness=1] (0.7,-0.26);
\end{tikzpicture}
\end{center}
\vspace{-0.cm}
\caption{Spacetime with cylindrical boundary. The entangling surface $\Sigma$ is a two-dimensional disk orthogonal to the boundary. Its boundary  $\cal P$ is a circle.}
\label{fig}
\end{figure}

\subsubsection{Heat kernel calculation for spacetime with cylindrical boundary}

The entanglement entropy is computed using the replica method,
\be
S(\Sigma) = (n\partial_n -1)W_s(n)|_{n=1}\,,
\lb{replica}
\ee
where $W_s(n)$ is the effective action of the field of spin $s$ considered on the Euclidean space $\mathcal{M}_{d,n}$ with a conical singularity at $\Sigma$. The effective action is expressed in terms of the heat kernel of the Laplace-type operator of the field $\triangle^{(s)}$,
\be
W_s(n) = -\frac{(-1)^{2s}}{2}\int_{\epsilon^2}\frac{d\tau}{\tau} \Tr K_n(\triangle^{(s)},\tau)\,.
\lb{Ws}
\ee
The heat kernel admits an asymptotic expansion for small proper time,
\be
\Tr K_n(\triangle^{(s)},\tau) \simeq \sum_{p=0} a_p(\triangle^{(s)},n) \,\tau^{(p-d)/2}\,, \qquad \tau\rightarrow 0\,.
\ee
in dimension $d$.
Since we are interested in the logarithmic part of the entanglement entropy in dimension $d=4$, we need to calculate the heat coefficient $a_4$ on $\mathcal{M}_{d,n}$.  For manifolds with boundaries the first few heat kernel coefficients are available, see \cite{Vassilevich:2003xt}. For the reader convenience we present the first three coefficients in Appendix. 

In order to fix the charge  $d$ in (\ref{slog}), we consider a particular case of a flat spacetime and choose the flat entangling surface $\Sigma$ to cross orthogonally the cylindrical boundary $\partial \mathcal{M}_4$. Then the replicated spacetime is the product of a two-dimensional cone $C_{2, n}$ with a disk $D_2 = \Sigma$, i.e. $\mathcal{M}_{4,n} = C_{2,n}\times \Sigma$. This geometrical setup simplifies the general structure of $s_{log}$  (\ref{slog}). Since the entangling surface $\Sigma$ crosses the boundary orthogonally then $F_e=0$.
Moreover,   $\Sigma$ is a flat disk so that $R_\Sigma=0$ and, since the spacetime is Minkowski, $W_{ijij}=0$. In the end, there remain in (\ref{slog}) only terms due to ${\cal P}$,  
\be
s_{log} = \frac{a-6d}{360\pi}\int_{\cal P} k_p \,,
\lb{slogF}
\ee 
where we used the fact that in the geometry under consideration one has $k_{\mu\nu}v^\mu v^\nu = k=k_p$ \;\;($\cal P$ is a circle of curvature $k_p$). 
Then our strategy is to directly compute the logarithmic term for this geometrical configuration for free fields of various spin. By comparing this term with (\ref{slogF}),
provided the $a$-charge is known, we deduce the value of the charge $d$. Below we provide details of this calculation.  The results are summarized  in Table \ref{tab4}.

\begin{table}[h]\renewcommand{\arraystretch}{1.5}
\begin{center}
\vspace{0.2cm}
\begin{tabular}{|c|c|c|c|c|c|}
\hline
Theory  & $a$ & $b$ & $c$ & $d$ & boundary condition \\
\hline
real scalar & $1$ & $1$ & $1$ & $1$ & Dirichlet \\
\hline
real scalar & $1$ & $1$ &  $\frac{7}{9}$ & -$\frac{2}{3}$ & Robin \\
\hline
Dirac spinor & $11$ & $6$ & $5$ & $1$ & mixed \\
\hline
gauge boson & $62$ & $12$ & $8$ & $7$ & absolute/relative \\
\hline
\end{tabular}
\vspace{-10pt}
\end{center}
\caption{Charges appearing in the conformal anomaly and in the logarithmic term in the entanglement entropy in dimension $d=4$.}
\label{tab4}
\end{table}

For scalar field with the Dirichlet condition, spin 1/2 field and the gauge boson the coefficient $d$ was computed in \cite{Fursaev:2013mxa}. For scalar and spinor fields
we just reproduce computations of \cite{Fursaev:2013mxa} (note that \cite{Fursaev:2013mxa}
used a slightly different normalization of $F_d$). For gauge fields our result differs
from \cite{Fursaev:2013mxa}. What was computed in \cite{Fursaev:2013mxa} is a 
sort of 'geometric' entropy of a gauge field. We use here an alternative method 
by trying to avoid unphysical properties of geometric entropy, see Sec. \ref{2met}.

\subsubsection{Scalar field: Dirichlet or Robin conditions}
We consider a real scalar field $\varphi$ with Dirichlet $(D)$ boundary condition
\be
\varphi|_{\partial\mathcal{M}} = 0\,,
\ee
or conformal Robin $(R)$ boundary condition
\be
\left(\nabla_N -  S\right)\varphi |_{\partial\mathcal{M}} = 0\,,
\lb{bc}
\ee
where $S=-\frac{1}{3}k$.  The Neumann  $(N)$ boundary condition corresponds to $S=0$ in (\ref{bc}).

The heat kernel on a product space $\mathcal{M}_{4,n} =C_{2,n}\times \Sigma$  factorizes as
\be  
K({\cal M}_{4,n}, \tau) = K(C_{2,n}, \tau)\times K (\Sigma, \tau)\,,
\lb{ker}
\ee
where $K(C_{2,n}, \tau)$ and $ K (\Sigma,\tau)$ are the heat kernels on $C_{2,n}$ and $\Sigma$, respectively. 
The heat kernel of a scalar Laplace operator on a flat cone $C_{2,n}$  is explicitly known,
\be
\Tr K(C_{2,n},\tau)=\frac{V_{C_{2,n}}}{4\pi \tau} + \frac{n}{12}(n^{-2}-1)\, .
\lb{Kcon}
\ee
The heat kernel on a two-dimensional disk $\Sigma=D_2$  is given by the asymptotic expansion
\be
 \Tr K (D_2,\tau) \hspace{-4pt} &\simeq& \hspace{-4pt}  \sum_{p=0} a_p(D_2) \,\tau^{(p-2)/2}\,,
\ee
where coefficients $a_p(D_2)\, , \, p=0\, , 1\, , 2$ are explicitly given in (\ref{a4}) (see \cite{Vassilevich:2003xt})  depending on the boundary conditions.
Using these formulae one finds 
\be
a^{(D)}_2(\Sigma) = a^{(N)}_2(\Sigma)=\frac{1}{12\pi}\int_{\cal P} k_p \,, \ \ \  a^{(R)}_2(\Sigma) = -\frac{1}{12\pi}\int_{\cal P} k_p
\lb{aa}
\ee
for respectively Dirichlet, Neumann and Robin boundary conditions.
Using the replica formula (\ref{replica}) we arrive at the logarithmic term
\be
s^{D(R)}_{log} = \mp \frac{1}{72\pi}\int_{\cal P} k_p \, ,
\lb{slogV0}
\ee
where one has $-$ for the Dirichlet boundary condition and $+$ for the conformal Robin boundary condition.

Now comparing (\ref{slogF}) with (\ref{slogV0}) we obtain the value of the charge $d$ for both boundary conditions (taking the value $a=1$ for the $a$-charge of a scalar),
\be
d =  \begin{cases} \displaystyle
\quad 1\,,&\quad {\rm for\,\, Dirichlet\ b.\, c.}\,,\\[2ex]  \displaystyle
-\frac{2}{3}\,,& \quad{\rm for\,\, Robin \ b\, . c.}\,.
 \end{cases}
\ee

\subsubsection{Spinor field}
Next we consider a Dirac field $\psi$ with mixed boundary conditions
\be
\Pi_-\psi|_{\partial\mathcal{M}} = 0\,,\qquad {\rm and} \qquad   \left(\nabla_N - S \right)\Pi_+\psi|_{\partial\mathcal{M}} = 0\,,
\ee
where $S=-\frac{1}{2}k\,\Pi_+$ and $\Pi_-$ and $\Pi_+=1-\Pi_-$ are two projectors on subsets of components of the field satisfying Dirichlet and Robin boundary conditions respectively (see \cite{Vassilevich:2003xt}). For the two-dimensional Laplacian $\triangle^{(1/2)}_\Sigma$,  acting on $2D$ spinors $\psi_2$, the corresponding boundary conditions are
\be
(\Pi_-)_2\psi_2|_{\cal P} = 0\,,\qquad {\rm and} \qquad \left(\nabla_N - S_2 \right)(\Pi_+)_2 \psi_2|_{\cal P} = 0\,,
\ee
with $S_2=-\frac{1}{2}k_p\,(\Pi_+)_2$ and $(\Pi_+)_2$, $(\Pi_-)_2$ the corresponding $2D$ projectors. 
Equations (\ref{ker}) and (\ref{a4}) still hold for the spinor operator and we have
\be
a^{(1/2)}_2(C_{2,n})  \hspace{-4pt}&=& \hspace{-4pt} -\frac{n}{12}(n^{-2}-1)\,,\\
a^{(1/2)}_2(\Sigma)  \hspace{-4pt}&=& \hspace{-4pt} \frac{\Tr_21-3\Tr_2(\Pi_+)_2}{12\pi}\int_{\cal P} k_p  \,.
\ee
One notices that $a^{(1/2)}_2(C_{2,n}) = -a^{(0)}_2(C_{2,n})$. This difference in the sign between the scalar and spinor fields is compensated by the same difference in sign in the effective action of the fields (see eq.(\ref{Ws})). Using the fact that $\Tr_2 I =2$ and $\Tr_2(\Pi_+)_2 =1$ we find for the logarithmic term 
\be
 s_{log} = \frac{1}{72\pi}\int_{\cal P} k_p \,.
\ee
Using the value $a=11$ for the $a$-charge of a Dirac fermion, this yields the value $d=1$.

\subsubsection{Gauge field}\label{gauge}

Our last example is a gauge field. Imposing the Lorentz gauge, one ends up with  two contributions to the partition function. One comes from the
vector field described by a vector Laplace operator $\Delta^{(1)}=-\nabla^2\delta^\mu_\nu +R^\mu_\nu$ and the other is a contribution of a ghost $c$
described by a scalar Laplace operator $\Delta^{(gh)}=-\nabla^2$. Then one finds for the heat kernel coefficient that
\be
a_4=a_4(\Delta^{(1)})-2a_4(\Delta^{(gh)})\, .
\lb{g1}
\ee
In the presence of boundaries one should impose boundary conditions both on the vector field and on the ghost.
There are two possible sets of boundary conditions (b.\,c.), the relative b.\,c.\, and the absolute b.\,c..
First, one introduces projectors $(\Pi_+)^\nu_\mu=\delta_\mu^\nu-N_\mu N^\nu$ and $(\Pi_-)_\mu^\nu=N_\mu N^\nu$.
The absolute b.\,c.~then is
\be
(\delta_\mu^\nu\nabla_N+k_\mu^\nu)V_\nu^+|_{\partial {\cal M}}=0\, ,  \quad
V^-_\mu |_{\partial{\cal M}}=0\, , \quad \partial_N c  |_{\partial{\cal M}}=0\,, 
\lb{g2}
\ee
while the relative b.\,c.~is
\be
(\nabla_N+k)V^-_\mu |_{\partial{\cal M}}=0\, , \quad V^+_\mu  |_{\partial{\cal M}}=0\, , \quad c  |_{\partial{\cal M}}=0\, .
\lb{g3}
\ee
In Minkowski spacetime the Ricci tensor $R_{\mu\nu}$ vanishes and the vector Laplace operator $\Delta^{(1)}$   reduces to 
the scalar Laplace operator acting on each component of the vector field $V^\mu$.  Then in our situation with a cylindrical boundary   it is convenient to use the cylindrical set of coordinates $(t,z,r,\phi)$, so that the boundary is at fixed $r$, and the angle $\phi$ is $2\pi$-periodic. In this coordinates system
only one component of the extrinsic curvature of $\partial{\cal M}_3$ is non-vanishing, $k_{\phi\phi}=r$ such that the trace $k=1/r$. Then the absolute b.\,c.~reduces to the following conditions
on the vector components
\be
\partial_r V^t  |_{\partial{\cal M}}=0\, , \quad \partial_r V^z  |_{\partial{\cal M}}=0\, ,  \quad V^r |_{\partial{\cal M}}=0\, , \quad (\partial_r+k) V^\phi  |_{\partial{\cal M}}=0\, , \quad \partial_r c  |_{\partial{\cal M}}=0\, .
\lb{g4}
\ee
On the other hand, the relative b.\,c.~reduces to
\be
 (\partial_r+k) V^r  |_{\partial{\cal M}}=0\, , \quad V^l  |_{\partial{\cal M}}=0\, , l=t,z,\phi \, , \quad c |_{\partial{\cal M}}=0\, .
\lb{g5}
\ee
Computing the heat kernel for each component, one uses the formula (\ref{ker}) for the product space ${\cal M}_{4,n}=C_{2,n}\times D_2$.
As we have seen, on a two-dimensional disk $D_2$ the Neumann and Dirichlet b.\,c.~lead to the same heat kernel coefficient $a_2$,  (\ref{aa}).
Therefore, the fields with the Dirichlet and Neumann b.\,c.~will have the same heat kernel coefficient on ${\cal M}_{4,n}$. 
For the component satisfying Robin b.\,c.~$(S=-k)$ one finds
\be
a_2^{(R)}(D_2)=-\frac{5}{12\pi}\int_{\cal P} k_{p}\, ,
\lb{g6}
\ee
where in this case $k_p=k$.
As we see from (\ref{g4}) and (\ref{g5}),  for both sets of boundary conditions one has one field satisfying Robin b.\,c.~and three fields satisfying Drichlet/Neumann b.\,c., and one ghost with Dirichlet/Neumann b.\,c..Thus, the total contribution to the heat kernel coefficient $a_4$  (\ref{g1}) in both cases will be the same.
Then for the absolute b.\,c., the total heat kernel $a_2$ on $D_2$ is a sum of all contributions,
\be
a_2(D_2)=a_2^{(D)}(D_2) +2a_2^{(N)}(D_2)+a^{(R)}(D_2)-2a_2^{(D)}(D_2)=-\frac{1}{3\pi}\int_{\cal P} k_{p}\, 
\lb{g7}
\ee
with the same answer for the relative b.\,c..

The other important point is that, as we already mentioned,  in flat spacetime a vector field is essentially a combination of 4 scalar fields
and thus one uses the result obtained for the heat kernel of a scalar field on a two-dimensional conical space $C_{2,n}$, (\ref{Kcon}). This means that the Kabat contact terms should not be
included in the heat kernel. This is the right way to compute the entanglement entropy of a vector field as   was advocated in \cite{Solodukhin:2015hma}.   It  leads to a positive area law in the entropy which occurs to be proportional to the number of physical degrees of freedom (2 in the case of a gauge field in 4 dimensions).
Putting things together one finds 
\be
s_{log}=\frac{1}{18\pi}\int_{\cal P}k_p\, 
\lb{g8}
\ee
for the logarithmic term due to the gauge field.  Comparing (\ref{g8}) with (\ref{slogF}) one can determine the value of the charge $d$ provided $a$ is known.
We do not calculate here the coefficient $a$ for gauge fields in (\ref{slog}). 
We simply take the $a$-charge in (\ref{slog}) the same as in the trace anomaly (\ref{d4}),
i.e. $a=62$ for the gauge field. This has two reasons. First, in the both formulas 
the coefficients stand at topological invariants and they coincide
for spin 0 and spin 1/2 fields. Second, $a=62$ is the only value which is consistent with the holographic calculation of both conformal anomaly and entanglement entropy.
With $a=62$ we find $d=7$.

\subsubsection{Remark on geometric entropy of gauge fields}\label{2met}

Field-theoretic computations of $a$-charge in the case of gauge 
fields may disagree with the value $a=62$. This disagreement was a matter of discussions
and different proposals, including the arguments that there is a 
contribution of some new degrees of freedom confined on the entangling surface, 
see \cite{Donnelly:2015hxa} for some references.

Since the $a$-charge stands at a topological term in the entropy, its 
value is sensitive to the presence of zero modes of  the corresponding Laplace operators.
These operators, like the Hodge-de Rham operator, appear in the effective
action when fixing the gauge. A careful quantization of gauge fields on spherical domains with conical singularities and mode-by-mode analysis  were done in \cite{DeNardo:1996kp}. The result of this
work is in agreement with the correct values $a=62$, $b=12$ in the log term of the entropy. 
In  ref.\cite{Fursaev:2013mxa} the calculations of \cite{DeNardo:1996kp} were used to derive 
value $d=2$ for the charge $d$. This value differs from what was found in the previous section since 
\cite{Fursaev:2013mxa} does not assume that the vector field is a combination of 4 scalars.

The method used in \cite{Fursaev:2013mxa},\cite{DeNardo:1996kp} implies 
that the entropy of gauge fields follows immediately from the corresponding effective action 
set on manifolds with conical singularities. One can call such entropy a geometric (or conical) entropy.
The drawback of the geometric entropy is that it has unphysical (negative) coefficient at the leading area term. Thus, although the geometric entropy yields $a=62$, it is not quite clear how it is related to the physical entanglement entropy, especially in the presence of boundaries. 

In subsequent calculations we use the results of section \ref{gauge} and the value $d=7$. 

\subsubsection{Relation between the charges}
Before going further we pause to discuss a possible relation between the charges $a\, , \ b\, , \ c\, , \ d$ that appear in the integral conformal anomaly and in the
logarithmic term in the entanglement entropy.  
Indeed, one may wonder whether they are independent or one of them may be reduced to a combination of others. 
A proper way to answer this question may involve the analysis of the correlation functions in  a generic conformal field theory 
on a manifold with boundaries. Such an analysis may be rather complicated and we are not inclined to dive into it in this paper. 
However, analyzing the different values of the charges collected in Table 2 we notice that, quite curiously,  there does exist
a linear relation between the charges: 
\be
d=\frac{1}{2}a-7b+\frac{15}{2}c\, . 
\lb{r1}
\ee
We note that the  existence  of this relation is not that trivial: assuming that $d$ is linearly expressed in terms of other charges as $d=\alpha a+ \beta b+ \gamma c$ 
one has to determine three constants $\alpha$, $\beta$ and $\gamma$ from four sets of values for  the charges available in Table 2.
This problem is clearly overdetermined and the fact that a solution to four constraints on three variables nevertheless exists is intriguing. It indicates that 
the $F_d$ term in the entanglement entropy (\ref{slog}) is likely to originate from all three terms in the conformal anomaly (\ref{d4}). Perhaps an analysis
in the spirit of  \cite{Fursaev:2016inw} and \cite{Fursaev:2013fta} would help to reveal the mechanism behind this ``empiric" relation.

\subsubsection{${\cal N}=4$ supermultiplet}

The ${\cal N}=4$ $SU(N)$ super-Yang-Mills (SYM) theory provides the simplest realization of AdS/CFT correspondence. The scale and Poincar\'e invariances of the theory combine into a larger symmetry -- the conformal invariance. The four-dimensional ${\cal N}=4$ $SU(N)$ SYM theory is thus a conformal field theory which admits a dual description in terms of $AdS_5$ supergravity. The field contents of this model are given by one gauge boson, two Dirac fermions and  six scalars, each field in the adjoint representation of $SU(N)$.
One is free to impose boundary conditions on the fields in the multiplet that normally breaks the supersymmetry. 
The maximal  amount of the supersymmetry  is preserved   provided one chooses the  Dirichlet b.\,c.~for a  half of the scalars and  and the 
 Robin b.\,c.~for  the other half. This 1/2-supersymmetric  configuration of fields is the most interesting to us since, as was discovered  in \cite{Astaneh:2017ghi}, in this case one finds a complete agreement between the anomaly calculation for the free fields  and the holographic calculation giving the anomaly in a strong coupling regime.

For the entire free ${\cal N}=4$ super-gauge multiplet, using Table \ref{tab4}, one finds for the values of the charges $a$, $b$, $c$, and $d$, 
\be
a &=& (N^2-1)\Big(3\cdot 1 + 3\cdot 1 + 2\cdot 11 + 1\cdot 62 \Big) \;=\; 90(N^2-1)\,,\lb{N4}\\[0.3ex]
b &=& (N^2-1)\Big(3\cdot 1 + 3\cdot 1 + 2\cdot 6 + 1\cdot 12 \Big) \;=\; 30(N^2-1)\,,\nonumber \\
c &=& (N^2-1)\Big(3\cdot 1 + 3\cdot \frac{7}{9} + 2\cdot 5 + 1\cdot 8 \Big) \;=\; \frac{70}{3}(N^2-1) \,,\nonumber \\
d &=& (N^2-1)\Big(3\cdot 1 - 3\cdot \frac{2}{3} + 2\cdot 1 + 1\cdot 7 \Big) \;=\; 10(N^2-1)\,.\nonumber 
\ee
The value of the charge $d$ is what concerns us here.
Putting things together, we obtain for the logarithmic term in the entanglement entropy  in the four-dimensional ${\cal N}=4$ $SU(N)$ SYM theory  (provided the  entangling surface $\Sigma$ crosses orthogonally the boundary $\partial {\mathcal{M}}_4$)  
\be
s_{log}^{(SYM)}=\frac{N^2-1}{8\pi}\left(  \Big[\int_\Sigma R_\Sigma+2\int_{{\cal P}=\partial\Sigma} k_p \Big]+\int_\Sigma \Tr \hat{k}_i^2 -2\int_{\cal P} \hat{k}_{\mu\nu}v^\mu v^\nu\right)\,.
\lb{EESYM}
\ee
This is our first main result. Later in the paper we shall reproduce this logarithmic term  holographically in the regime of large $N$.

Provided one uses the geometric (conical) entropy for the  gauge fields (which predicts $d=2$)  one finds the expression as above  but without the factor 2 in the last term
due to the extrinsic curvature. As we show in the next section this is not in agreement with the holographic calculation.

\section{Holographic calculation}
In this part of the paper  we would like to perform the holographic calculation of the entanglement entropy in the presence of boundaries.
In order to arrive at an appropriate prescription for this holographic computation, we need first to know what is the holographic dual of a boundary conformal field theory (BCFT).  In what follows we first review such a duality in the context of the AdS/BCFT correspondence and then introduce our holographic proposal for the calculation of the entanglement entropy. 

\subsection{General prescription}
The geometrical set-up on which we construct our holographic picture is as follows. The $(d + 1)$-
dimensional anti-de Sitter spacetime $AdS_{d+1}$ has $\CM_d$ as a conformal boundary. We consider
a situation where $\CM_d$ has a boundary by itself, we denote it by $\p\CM_d$. This boundary, being extended to the bulk, forms a $d$-dimensional
surface $\CS_d$. Thus, the boundary of the bulk spacetime has two components, $\CS_d$ and $\CM_d$; these two meet at a common boundary such that $\p\CS_d=\p\CM_d$.

Constructing the dual theory, one first has to prescribe how $\p\CM_d$  is extended into the bulk. Here we briefly review two available prescriptions for constructing the holographic dual of a boundary conformal field theory.

\vspace{0.5cm}
\noindent{\it{Takayanagi's prescription.}}
In this prescription the gravitational action consists of the Einstein-Hilbert action to which is added a Gibbons-Hawking term as well as a boundary cosmological constant on the holographic boundary $\CS_d$,
\be
W_{grav}^{T}=-\frac{1}{16\pi G}\int_{AdS_{d+1}}(R+2\Lambda)-\frac{1}{8\pi G}\left[\int_{\CM_d}K+\int_{\CS_d}(K+T)\right]\, ,
\ee
where $K$ denotes the trace of the extrinsic curvature tensors on the boundaries, $\CM_d$ and $\CS_d$. Varying the action with respect to the boundary metric, $\gamma_{ij}$, one can determine the profile of $\CS_d$. Doing so one arrives at 
\be
K_{ij}-\gamma_{ij}(K+T)=0\,, 
\lb{Kij}
\ee 
which should be solved perturbatively in the radial direction of the bulk in order to obtain the shape of $\CS_d$. These equations are solvable only in some special cases where additional symmetries are present. 

\vspace{0.5cm}
\noindent{\it{Restricted Takayanagi's prescription.}}
In more general situations one restricts   \cite{SJ}  the tensorial equations (\ref{Kij}) to a single scalar type equation imposed on the trace of the extrinsic curvature of the boundary, 
\be
K=-\frac{d}{d-1}T\, .
\lb{K}
\ee
A solution to this equation should give us the shape of ${\cal S}_d$. Notice however that equation (\ref{K}) alone does not follow from any variational principle.
In what follows, for the sake of simplicity, we introduce a new parameter $m$ and rewrite the tension as $T=(d-1)\tanh(m)$.

\vspace{0.5cm}
\noindent{\it Minimal surface prescription.}
In an alternative prescription \cite{Astaneh:2017ghi}  the boundary $\CM_d$ is minimally extended into the bulk, i.e.
\be
K=0 \  \ \ \text {on}\  \ \ \CS_d\, .
\ee
This equation does follow from a variational principle provided one adds a volume functional on $\CS_d$ to the gravitational action,
\be
W_{grav}^{min}=-\frac{1}{16\pi G}\int_{AdS_{d+1}}(R+2\Lambda)-\frac{1}{8\pi G}\left[\int_{\CM_d}K+\int_{\CS_d}\lambda\right]\, ,
\ee
where the induced metric on $\CS_d$ is given in terms of embedding functions.
As was found in \cite{Astaneh:2017ghi}, this particular choice of profile of the holographic boundary guaranties 1/2 supersymmetry preservation in the boundary theory.

We are now ready to investigate how the holographic entanglement entropy is modified in the presence of boundaries. The entanglement entropy is defined in a conformal field theory living on $\CM_d$ for an
entangling surface $\Sigma$ which lies on a hypersurface of constant time. We consider a situation
when $\Sigma$ intersects the boundary $\partial\CM_d$ so that the intersection is a $(d-3)$-dimensional surface $\CP$ (which may have several components). In the original proposal by Ryu and Takanayagi  \cite{Ryu:2006bv}
one considers a minimal surface $\CH$ which bounds $\Sigma$ and goes into the bulk of the anti-de
Sitter spacetime. In the present set-up the hypersurface $\CH$ has boundary $\partial\CH$ which itself
bounds the surface $\CP$. Although other prescriptions may be possible, we extend ${\cal H}$ all the way till the boundary ${\cal S}_d$ so that
$\partial {\cal H}={\cal H}\cap {\cal S}_d$.  As a result, the holographic entanglement entropy which is governed by the area of the minimal surface $\CH$, 
\be
S_{hol}[\Sigma,\CP]=\frac{A[\CH]}{4G}\, ,
\ee
will carry some information about the geometry of the hypersurface at which $\Sigma$ intersects $\p\CM_d$, i.e. $\CP$.
This is of course not the most general prescription. 
One can also consider a modification of RT proposal and add the area of the boundary of the minimal surface, 
\be
S_{hol}[\Sigma,\CP]=\frac{1}{4G}(A[\CH]+\eta A[\p\CH])\, ,
\ee
where $\eta$ is a new constant parameter which should be determined independently. In the following sections, we will see how this proposal works for a boundary conformal field theory in three and four dimensions and make comparison with what we have obtained on the field theory side. Our principal goal here is to identify the contributions to the
logarithmic term in the entropy  that are due to the boundary ${\cal P}$ of the entangling surface $\Sigma$. They should be then combined with the contributions due to the 
bulk of $\Sigma$ that are 
already well studied in the literature,  see \cite{Solodukhin:2008dh}, \cite{Myers:2013lva}.

\subsection{Form of the AdS metric}

We shall cast the AdS$_{d+1}$ bulk metric in the form 
\be
\quad ds^2=\frac{d\rho^2}{4\rho^2}+\frac{1}{\rho}\left(-dt^2+dr^2+(\gamma_{ij}-k_{ij}r)^2dx^idx^j\right)\,, \qquad i,j=1,2,\cdots ,d-2\, .
\lb{gk}
\ee
where
\be
(\gamma_{ij}-k_{ij}r)^2=(\gamma_{i\ell}-k_{i\ell}r)(\gamma^\ell_j-k^\ell_jr)=\gamma_{ij}-2k_{ij}r+(k^2)_{ij}r^2\, .
\lb{kk}
\ee
The boundary ${\cal M}_d$ is defined by $r=0$ so that $\gamma_{ij}$ is the metric on $\partial{\cal M}_d$. 
The only non-vanishing component of the outward normal vector is $N^r=-1$  so that $k_{ij}$ in (\ref{gk}) and (\ref{kk}) coincide with the
$ij$ components of the extrinsic curvature of $\partial {\cal M}_d$
defined with respect to $N^\mu$. Other, the $tt$ and $it$ components of the extrinsic curvature
are equal to zero.

Using a normal coordinate system on $\partial{\cal M}_d$
one has
\be
\gamma_{ij}(x)=\delta_{ij}+O(x^2)\,.
\lb{gam}
\ee
The contribution which we are looking for is at most linear in the extrinsic curvature $k_{ij}$ so that we can neglect all terms quadratic in $x$ in (\ref{gam}) as well as
the higher order terms in $r$ which may be present in (\ref{gk}). 

\subsection{Calculation in $d=3$}
In three dimensions the metric (\ref{gk}) takes the simple form
\be
ds^2=\frac{d\rho^2}{4\rho^2}+\frac{1}{\rho}(-dt^2+dr^2+(1-k_{11}r)^2dx^2_1)\, .
\ee
In this set-up the entangling surface is defined as
\be
\Sigma \ : \ t=0\,, \;\; x_1=0\,,
\lb{ES}
\ee
and its extension into the bulk is specified as
\be
\CH \ : \ t=0\,, \;\; x_1=x(r,\rho)\,.
\ee
The area of the surface $\CH$ can be obtained as
\be\label{A[H3]}
A[\CH]=\int_{\epsilon^2}\frac{d\rho}{2\rho^{3/2}}\int_{r(\rho)}dr\sqrt{1+(4\rho x'^2_\rho+x'^2_r)(1-k_{11}r)^2}\, ,
\ee
where $x'_\rho=\p_\rho x(r,\rho)$ and $x'_r=\partial_r x(r,\rho)$.

Minimizing the area functional (\ref{A[H3]}), subject to condition $x(r,\rho=0)=0$, we find that the following asymptotic profile suitably solves the Euler-Lagrange equation,
\be
x(r,\rho)=\rho^{3/2}\left(c_1+c_2 r+c_2 k_{11}r^2+\CO(r^3)\right)+\cdots\, ,
\lb{xro}
\ee
where $c_1$ and $c_2$ are two constants. 
Then we substitute the solution $x(r,\rho)$ in (\ref{A[H3]}) and integrate over $r$ first. Curiously, the logarithmic term in (\ref{A[H3]}) does not depend on
the exact profile   (\ref{xro}). 
However, we do need to know the profile of $\CS_3$ that appears in the lower limit of the integration over $r$ in (\ref{A[H3]}).
In \cite{Astaneh:2017ghi}  it was found to be 
\be\label{r(rho)3}
r(\rho)=r_0\rho^{1/2}+r_1\rho+r_2\rho^{3/2}+\cdots \, ,
\ee
where
\be
r_0=\sinh(m)\, ,  \  \  \ r_1=-\frac{\alpha_1}{4}\cosh^2(m)\, , \ \ \  r_2=\frac{1}{24}\sinh(m)\cosh^2(m)(7\alpha_1^2-16\alpha^2)\,, 
\lb{d3-1}
\ee
and
\be
\alpha_1=-k\ , \ \alpha_2=\frac{1}{2}(k^2-\Tr k^2)\, .
\ee
Only the first term in  (\ref{r(rho)3}) contributes to the logarithmic term,
\be
A_{log}[\CH]=\sinh(m)\ln\epsilon\, .
\ee
Then we find for the logarithmic term in the holographic entropy
\be
S_{log}=\frac{A_{log}[\CH]}{4G}=\frac{1}{4G}\sinh(m)\ln\epsilon\, .
\lb{shol}
\ee
Obviously, this term vanishes if the boundary ${\cal S}_3$ is minimal.

There may also be some contribution from the boundary of the holographic surface. To find it we set
\be
\p\CH:\ t=0\,, \;\; x_1=x(\rho) \equiv  x(\rho, r(\rho))\, ,
\ee
where $x(\rho, r)$ is given by (\ref{xro}) and $r(\rho)$ by \eqref{r(rho)3},
in the induced metric on $\CS_3$,
\be
ds^2_{\p\CH}=\frac{1}{4\rho^2}\left(1+4\rho\, r'^2(\rho)+4\rho(1-k_{11}r(\rho))^2x'^2_\rho\right)d\rho^2\, ,
\ee
and the area reads
\be
A[\p\CH]=\int_{\epsilon^2}\frac{d\rho}{2\rho}\sqrt{1+4\rho\, r'^2(\rho)+4\rho(1-k_{11}r(\rho))^2x'^2_\rho}\,.
\ee
Again, the logarithmic term does not depend on the exact profile (\ref{xro}) and we find
\be
A_{log}[\p\CH]=-\cosh(m)\ln\epsilon\, .
\ee
The modified holographic prescription then yields
\be
S'_{log}=\frac{1}{4G}\left(\sinh(m)-\eta \cosh(m)\right)\ln\epsilon
\lb{S'}
\ee
for the logarithmic part of the entropy. If there are several intersections of the entangling surface $\Sigma$ with the boundary $\partial{\cal M}_3$ then
(\ref{shol})   and (\ref{S'}) would be proportional to the number of these intersections. This is in agreement with the calculations
in  $d=3$  on the field theory side.

\subsection{Calculation in $d=4$}
The computation of the entanglement entropy in four dimensions follows the same path as that of in three dimensions.  We start with the metric of AdS$_5$
\be
ds^2=\frac{d\rho^2}{4\rho^2}+\frac{1}{\rho}\left(-dt^2+dr^2+(\gamma_{ij}-k_{ij}r)^2dx^idx^j\right)\,, \qquad i=1,2\, .
\ee
The entangling surface is given by (\ref{ES}). Coordinate $x_2$ is thus the coordinate along ${\cal P}$, the intersection of $\Sigma$ with $\partial{\cal M}_4$. 
The area $\CH$ is found to be
\be\label{A[H4]}
A[\CH]=\int_{\epsilon^2}\frac{d\rho}{2\rho^2}\int_{r(\rho)}dr\sqrt{1+(4\rho x'^2_\rho+x'^2_r) (1-k_{11}r)^2}\, (1-k_{22}r)\, .
\ee
In four dimensions a suitable profile $x=x(r,\rho)$, subject to boundary condition $x(r,\rho=0)=0$, which  minimizes the area functional above is 
\be
x(r,\rho)=\rho^2 \left(c_1+c_2r+c_2(k_{11}+\frac{k_{22}}{2})r^2+\CO(r^3)\right)+\cdots\, .
\lb{xxx}
\ee
Similarly to what we did in three dimensions, we substitute this profile  in  equation \eqref{A[H4]} and perform the integration  over $r$ which is bounded from below by the intersection with $\CS_4$. In the same way as in three dimensions the logarithmic term in the area does not depend on the exact form (\ref{xxx}). However, it does depend on the function $r(\rho)$ standing in the lower limit in the integral over $r$.  This function determines the position of the boundary ${\cal S}_4$. Its exact form has been found in \cite{Astaneh:2017ghi},
\be\label{r(rho)}
r(\rho)=r_0\rho^{1/2}+r_1\rho+r_2\rho^{3/2}+r_3\rho^2+\cdots \, ,
\lb{solution}
\ee
where
\be
&&r_0=\sinh(m)\, ,\nonumber\\
&&r_1=-\frac{1}{6}\alpha_1\cosh^2(m)\, ,\nonumber\\
&&r_2=\frac{1}{6}\sinh(m)\cosh^2(m)(\alpha_1^2-2\alpha_2)\, ,\nonumber\\
&&r_3=\frac{1}{216}\Big(\cosh^4(m)\left(-47\alpha_1^3+144\alpha_1\alpha_2-162\alpha_3\right)\nonumber\\
&&\qquad+2\cosh^2(m)\left(20\alpha_1^3-63\alpha_1\alpha_2+81\alpha_3\right)\Big)\, ,
\ee
and
\be
&&\alpha_1=-k\, , \nonumber\\
&&\alpha_2=\frac{1}{2}(k^2-\Tr k^2)\, ,\nonumber\\
&&\alpha_3=-\frac{1}{6}(k^3-3k\Tr k^2+2\Tr k^3)\, .
\ee
Only the first two terms in (\ref{solution}) contribute to the logarithmic term in the area.
Then performing the integral over $\rho$ one can extract the logarithmic part of $A[\cal{H}]$, i.e.
\begin{eqnarray}
A_{log}[\CH] &=& \int dx_2(r_1-\frac{1}{2}k_{22}r_0^2)\ln\epsilon\nonumber  \\
&=&\int dx_2 \, \frac{1}{2}\left( -(k_{22}-\frac{1}{3}k)\cosh^2(m)+k_{22}\right)\ln\epsilon\, .
\label{1}
\end{eqnarray}
The contribution due to $\CP=\partial{\cal M}_4  \cap \Sigma= \partial\Sigma$ in the logarithmic part of the holographic entanglement entropy is then
\be
S^{(hol)}_{log}[\CP]=\frac{A_{log}[{\CH}]}{4G}=\frac{N^2}{4\pi} \left(-\int_\CP(k_{22}-\frac{1}{3}k)\cosh^2(m)+\int_\CP k_{22}\right)\ln\epsilon\,,
\lb{SholoP}
\ee
where we defined $N^2= \pi/(2G)$ according to the AdS/CFT dictionary. Equation (\ref{SholoP}) should be combined with the part of the holographic entanglement entropy due to
the bulk of the  entangling surface $\Sigma$, see  \cite{Solodukhin:2008dh}, \cite{Myers:2013lva},
\be
S^{(hol)}_{log}[\Sigma]=\frac{N^2}{8\pi}\left( \int_\Sigma R_\Sigma+\int_\Sigma \Tr \hat{k}^2_i\right)\ln\epsilon\, .
\ee
Altogether, the holographic entropy becomes
\be
S^{(hol)}_{log}[\Sigma,\CP]=\frac{N^2}{8\pi}\left( \Big[\int_\Sigma R_\Sigma+2\int_{\CP=\partial\Sigma} k_p\Big]+\int_\Sigma \Tr \hat{k}^2_i -2\cosh^2(m)\int_\CP \hat{k}_{ij}v^i v^j\right)\ln\epsilon\, ,
\lb{S}
\ee
where $k_p$ is the extrinsic curvature of $\CP$ inside $\Sigma$. In our case $k_p=k_{22}$  and $\hat{k}_{ij}v^i v^j=k_{22}-\frac{1}{3}k$, $v^i$ is a tangent vector to $\CP$ with the only non-vanishing component  $v^2=1$. The first term in (\ref{S}) is the Euler number of $\Sigma$.

\medskip

Now we can compare this holographic result with the field theory calculation (\ref{EESYM}) for the ${\cal N}=4$ super-gauge multiplet. We see that the two calculations are identical
for large $N$ provided the boundary ${\cal S}_4$ is a minimal surface $(m=0)$. This is our second main result. It shows that the agreement between the field theory calculation and the
holographic calculation, already established for the conformal anomaly in   \cite{Astaneh:2017ghi},  is extended to the entanglement entropy.

\medskip

Having in mind the possible generalization (\ref{S'}) of the holographic entropy in three dimensions, we can also compute the area of the boundary of the holographic surface $\cal H$.
Its area in dimension $d=4$ reads
\be
A[\p\CH]=\int_{\epsilon^2}\frac{d\rho }{2\rho^{3/2}}\sqrt{1+4\rho r'^2(\rho)+4\rho(1-k_{11}r(\rho))^2x'^2_\rho}(1-k_{22}r(\rho))\, ,
\ee
where $r(\rho)$ is given by (\ref{solution}) and $x(\rho)$ is obtained by substituting $r=r(\rho)$ into (\ref{xxx}).
The logarithmic term in the area, once again, does not depend on the exact profile of $x(\rho)$  and is found to be
\be
A_{log}[\p\CH] &=& \int dx_2\left(r_0\sqrt{1+r_0^2}k_{22}-\frac{2r_0r_1}{\sqrt{1+r_0^2}}\right)\ln\epsilon \nonumber \\
&=&\sinh(m)\cosh(m)\int dx_2(k_{22}-\frac{1}{3}k)\ln\epsilon\, ,
\ee
which can be written as
\be
A_{log}[\p\CH]=\sinh(m)\cosh(m)\int_{\cal P}\hat{k}_{ij}v^iv^j\ln\epsilon\, .
\lb{al}
\ee
We see that this term vanishes if the boundary ${\cal S}_4$ is minimal. In  the non-minimal case $(m\neq 0)$, by adding  (\ref{al}) to (\ref{1}) as in (\ref{S'})
one can change the coefficient in front of the invariant $F_e$ in the entropy.  Validity of this generalization needs to be further investigated.

\section{Conclusions}
In this paper we have performed the explicit, both field theoretic and  holographic, calculations of the entanglement entropy in the situation where the entangling surface $\Sigma$ crosses  orthogonally the boundary  $\partial{\cal M}_d$ of the spacetime ${\cal M}_d$. In this case the entangling surface has a boundary $\cal P$ and we focused on the contributions to the logarithmic term in the entropy
that are due to $\cal P$. 

We have found that one contribution, proportional to the extrinsic curvature of $\cal P$ is precisely of the form which,
being combined with the bulk of $\Sigma$,  produces the topological Euler number of $\Sigma$. The other  contribution, due to the traceless part of the extrinsic curvature of $\partial{\cal M}_d$, takes precisely the form predicted in \cite{Fursaev:2013mxa}. If one uses the prescription of section  
\ref{gauge} to compute the entanglement entropy for the gauge fields, the field theoretical computation for the free ${\cal N}=4$ super-gauge multiplet (\ref{EESYM}) and the holographic computation are in complete agreement. This is  provided that on the field theory side one chooses the boundary conditions which preserve $1/2$ of the supersymmetry
and that on the gravity side the boundary ${\cal S}_d$ is defined to be a minimal surface. This finding extends the earlier result obtained in \cite{Astaneh:2017ghi}
on the agreement between the two calculations of the conformal anomaly in the presence of boundaries.  
On the other hand, if one uses the geometric (or conical) entropy of gauge fields the result disagrees with the above holographic prescription due to a coefficient at the term with the traceless part of the extrinsic curvature.

We have considered several generalizations of the holographic calculation.
In particular, we have obtained the holographic result in the case when boundary ${\cal S}_d$ has  a non-vanishing constant extrinsic curvature.
As another generalization, we also have considered the possibility of adding a contribution of the area of the boundary $\partial {\cal H}$ of the holographic 
surface $\cal H$ to the holographic formula for the entropy. Validity of these generalizations needs to be further examined. 
In principle, one could use this freedom  to adjust the parameters in these generalizations to achieve 
the agreement with the field theoretic computations where the geometric entropy is used in the gauge sector.

On the other hand, as we demonstrate it  in this paper,
 the minimal surface prescription  of \cite{Astaneh:2017ghi} for the boundary ${\cal S}_d$, the standard Ruy-Takayanagi  holographic formula  \cite{Ryu:2006bv} 
 and the physically motivated way to compute the entropy due to gauge fields is all one needs in order to have complete agreement between the two sides of the holographic duality.

\bigskip

\noindent {\bf Acknowledgements.} The work of A.F.A is supported by Iran National Science Foundation (INSF).

\appendix
\section{Heat kernel coefficients}
\setcounter{equation}0
\numberwithin{equation}{section}

In this section we present results collected in \cite{Vassilevich:2003xt} that we make use of in this paper.

Let $\mathcal{M}$ be a smooth Riemannian manifold of dimension $d$, with a smooth boundary $\partial\mathcal{M}$. A differential Laplace type operator $\triangle=-\nabla^2+E$ acts on a vector bundle $V$ over $\mathcal{M}$. The heat kernel of $\triangle$ is defined as
\be
\Tr K(\triangle, \tau) = \Tr e^{-\tau\triangle}\,,
\lb{a1}
\ee
and admits an asymptotic expansion for small proper time,
\be
\Tr K(\triangle,\tau) \simeq \sum_{p=0} a_p(\triangle) \,\tau^{(p-d)/2}\,, \qquad \tau\rightarrow 0\,.
\lb{a2}
\ee

We consider the case of manifolds with boundaries, and fields $\phi$ with mixed boundary conditions
\be
\Pi_-\phi|_{\partial \mathcal{M}} = 0\,,\qquad {\rm and} \qquad   \left(\nabla_N - S \right)\Pi_+\phi|_{\partial \mathcal{M}} = 0\,,
\lb{a3}
\ee
where $\Pi_-$ and $\Pi_+=1-\Pi_-$ are two projectors on subsets of components of the field satisfying Dirichlet and Robin boundary conditions respectively. The heat coefficients $a_p$ up to $p=2$ are listed below:
\be
a_0 &=& \frac{1}{(4\pi)^{d/2}}\int_{\mathcal{M}} \tr_V(I)\,,\\
a_1 &=& \frac{1}{4(4\pi)^{(d-1)/2}}\int_{\partial\mathcal{M}} \tr_V(\chi)\,,\\
a_2 &=& \frac{1}{6(4\pi)^{d/2}}\left[\int_{\mathcal{M}} \tr_V(R - 6E) +2 \int_{\partial\mathcal{M}} \tr_V(k + 6S) \right]\,,
\lb{a4}
\ee
where $\chi = \Pi_+ -\Pi_-$, and $R$ is the Ricci scalar of $\cal{M}$.
Dirichlet and Robin boundary conditions are recovered for $\Pi_+ = 0$ and $\Pi_- = 0$, respectively.

\end{document}